\documentclass[aps,prx,superscriptaddress,reprint,showpacs,floatfix]{revtex4-1}
\usepackage{graphicx}
\usepackage{dcolumn}
\usepackage{bm}
\usepackage{xcolor}
\usepackage{relsize}
\usepackage{amsmath}
\usepackage{braket}
\usepackage{gensymb}
\usepackage{hyperref}

\begin{document}

\preprint{APS/123-QED}

\title{Polariton pattern formation and photon statistics of the associated emission}

\author{C. E. Whittaker}
\email{cewhittaker1@sheffield.ac.uk}
\affiliation{Department of Physics and Astronomy, University of Sheffield, Sheffield S3 7RH, United Kingdom}%
\author{B. Dzurnak}
\affiliation{Department of Physics and Astronomy, University of Sheffield, Sheffield S3 7RH, United Kingdom}%
\author{O. A. Egorov}%
\affiliation{Technische Physik, Wilhelm-Conrad-R{\"o}ntgen-Research Center for Complex Material Systems, Universit{\"a}t W{\"u}rzburg,
Am Hubland, D-97074, W{\"u}rzburg, Germany}%
\author{G. Buonaiuto}
\affiliation{Department of Physics and Astronomy, University of Sheffield, Sheffield S3 7RH, United Kingdom}%
\author{P. M. Walker}
\affiliation{Department of Physics and Astronomy, University of Sheffield, Sheffield S3 7RH, United Kingdom}%
\author{E. Cancellieri}
\affiliation{Department of Physics and Astronomy, University of Sheffield, Sheffield S3 7RH, United Kingdom}%
\author{D. M. Whittaker}
\affiliation{Department of Physics and Astronomy, University of Sheffield, Sheffield S3 7RH, United Kingdom}%
\author{E. Clarke}
\affiliation{EPSRC National Centre for III-V Technologies, University of Sheffield, Sheffield S1 3JD, United Kingdom}
\author{S. S. Gavrilov}
\affiliation{Institute of Solid State Physics, RAS, Chernogolovka 142432, Russia}
\author{M. S. Skolnick}
\affiliation{Department of Physics and Astronomy, University of Sheffield, Sheffield S3 7RH, United Kingdom}%
\author{D. N. Krizhanovskii}
\email{d.krizhanovskii@sheffield.ac.uk}
\affiliation{Department of Physics and Astronomy, University of Sheffield, Sheffield S3 7RH, United Kingdom}%

\date{\today}

\begin{abstract}
We report on the formation of a diverse family of transverse spatial polygon patterns in a microcavity polariton fluid under coherent driving by a blue-detuned pump. Patterns emerge spontaneously as a result of energy-degenerate polariton-polariton scattering from the pump state to interfering high order vortex and antivortex modes, breaking azimuthal symmetry. The interplay between a multimode parametric instability and intrinsic optical bistability leads to a sharp spike in the value of second order coherence $g^{(2)}(0)$ of the emitted light, which we attribute to the strongly superlinear kinetics of the underlying scattering processes driving the formation of patterns. We show numerically by means of a linear stability analysis how the growth of parametric instabilities in our system can lead to spontaneous symmetry breaking, predicting the formation and competition of different pattern states in good agreement with experimental observations.

\end{abstract}

\pacs{Valid PACS appear here}

\maketitle

\section{\label{sec:level1}Introduction}

The phenomenon of spontaneous pattern formation occurs ubiquitously in science in a diverse range of nonlinear extended media \cite{karsenti,Lehn2400,Zeng1302,cross,Arecchi1999}. The universal mechanism by which stationary patterns emerge from an initially symmetric state has fundamental conceptual significance, combining a cross-talk mechanism acting on different points in space and local nonlinear interactions \cite{Turing37}. As the system is taken out of equilibrium by varying some control parameter, the spontaneous growth of new components via interactions allows localized structures balanced by propagation and nonlinearity to form. In optics, diffraction and nonlinear mixing of electromagnetic waves allows transverse localized structures in various nonlinear media \cite{cross,Arecchi1999}.

On the other hand, topological entities such as quantized vortices characterized by a phase winding around a core also play an important role in pattern formation in many areas of science, including superconductors and stirred atomic Bose-Einstein condensates (BECs), where patterns formed by vortices with a single winding $m=1$  were addressed  \cite{ESSMANN1967526,doi:10.1063/1.1656923,PhysRevLett.96.077005,PhysRevLett.84.806,Abo-Shaeer476}, whereas higher order vortices are unstable. Similarly, in optical systems, high order vortex beams have long been known to be less stable than single vortices \cite{Nye165,doi:10.1080/09500349214551011,BASISTIY1993422}, and the experimental preparation of such beams and nonlinear conversion between them is challenging \cite{PhysRevLett.114.173901, ApurvChaitanya:16}, remaining an obstacle towards realizing many useful applications in optical information processing \cite{Molina-Terriza2007} and quantum entanglement \cite{Mair2001}. One intriguing question is whether the nonlinear wave mixing processes which cause generation of new translational momentum components and drive pattern formation in optical systems \cite{cross,Arecchi1999}, may lead to the efficient nonlinear generation of vortex beams with sizable orbital angular momentum (OAM). In this work we address such instabilities and the new class of spontaneous patterns that result.

At the same time, in a whole host of nonlinear optical systems, both conservative (optical fibers,  waveguides, etc.) and open-dissipative with gain and loss (fiber parametric amplifiers, lasers, nonlinear optical cavities, etc.), the interplay between nonlinearity (of or similar to the kind responsible for pattern formation) and temporal dispersion or spatial diffraction has been shown to lead to enhanced fluctuations of optical fields in space and time and even the emergence of extreme-value statistics characteristic of the formation of rogue waves \cite{Solli2007,Dudley2014}. Nevertheless, the classical nature of pattern formation means that it is usually studied without consideration of the quantum nature of light. Studies of quantum effects in the context of pattern formation have discussed quantum entanglement, squeezing and spatial correlations \cite{kolobov2007quantum}.

Here we study a nonequilibrium fluid of exciton-polaritons  in a semiconductor microcavity, which offer a rich platform for studying pattern formation, owing to the combination of diffractive spatial propagation and giant repulsive exciton-mediated nonlinearity \cite{paulncom}. Furthermore, direct access to the photonic component of the intracavity field allows quantum statistical properties to be probed. Importantly, the bosonic nature of polaritons allows the formation of BECs under non-resonant pumping \cite{Kasprzak2006} and also kinetic condensates in the optical parametric oscillator (OPO) regime, exhibiting increased temporal coherence \cite{PhysRevLett.97.097402} and long-range order similar to BECs \cite{PhysRevB.72.125335,PhysRevLett.96.176401}. The spontaneous U(1) gauge symmetry breaking also implies the appearance of Goldstone modes \cite{PhysRevA.76.043807} and the Berezinskii-Kosterlitz-Thouless mechanism in the OPO condensates \cite{PhysRevX.5.041028}. Polaritons evolve according to the Gross-Pitaevskii equation (GPE), which also describes many other condensed matter systems. A significant consequence is quantized vorticity, which emerges from certain solutions of the GPE.  So far single vortices have been observed in polariton BECs, resonant and OPO superfluids \cite{Lagoudakis2008,Roumpos2011a,Boulier2015,Anton:12}. However, complex patterns driven by higher order vortex modes have not been investigated. Theoretical studies of pattern formation in polariton systems have demonstrated the possibility of roll, labyrinthine and honeycomb patterns \cite{PhysRevB.89.245307}. Experimentally, standing wave patterns have been reported using annular pumping geometries \cite{manni,sun} in nonresonantly pumped BECs and an extended triangular pattern was observed in a double cavity structure \cite{ardizzone}. Statistical properties of the emission of polariton patterns remain unexplored. 

In this paper we study pattern formation in an unconfined polariton fluid in a single monolithic cavity. Under coherent driving by an external Gaussian pump, strong $\chi^{(3)}$ nonlinearity causes previously unreported spontaneous generation of high order vortex beams with high efficiency, resulting in a diverse family of patterns, including dipoles, quadrupoles, pentagons, hexagons, heptagons and octagons. This new class of patterns arises due to a novel form of highly efficient energy-degenerate parametric scattering which produces integer OAM components with opposite polarity, similar to a recent theoretical proposal \cite{1704.00397}. Interestingly, it allows polygons with odd numbers of vertices, which are to the best of our knowledge previously unobserved in other pattern forming systems, relying on the unique and novel pattern forming mechanism in our system. 

We also investigate statistical properties of the emission (intensity noise and second order coherence) governed by the interplay between the multi-mode ``pattern-forming'' parametric instability and bistability. This latter feature is induced by the blue-detuning of our pump with respect to the lower polariton branch (LPB) ground state \cite{baas}, which creates favourable conditions for the formation of polygon patterns. In single mode cavities with bistability superbunching and antibunching are predicted when the strength of interactions between two polaritons is comparable to the decay rate \cite{Drummond1999,PhysRevB.73.193306}. We show that, despite the interaction strength between individual polaritons in a pumped area being orders of magnitude less than the decay rate, a strong bunching is observed at the bistable transition. Such an effect arises from the strongly superlinear temporal dynamics of the total polariton field (so-called ``blowup'' \cite{PhysRevB.90.205303,PhysRevB.92.205312}) associated with the accumulation of energy in multiple scattered modes due to parametric instabilities, which leads to strong amplification of quantum fluctuations. 

The article is organized as follows: In Sec. II, we describe the sample and excitation scheme, present data showing the genesis and evolution of patterns with increasing pump power and offer a phenomenological interpretation of the results. In Sec. III, we investigate the photon statistics of the microcavity emission. In Sec IV, we perform numerical calculations including a full linear stability analysis. We provide concluding remarks in Sec. V.

\section{Pattern formation}

\begin{figure}
\center
\includegraphics[scale=1]{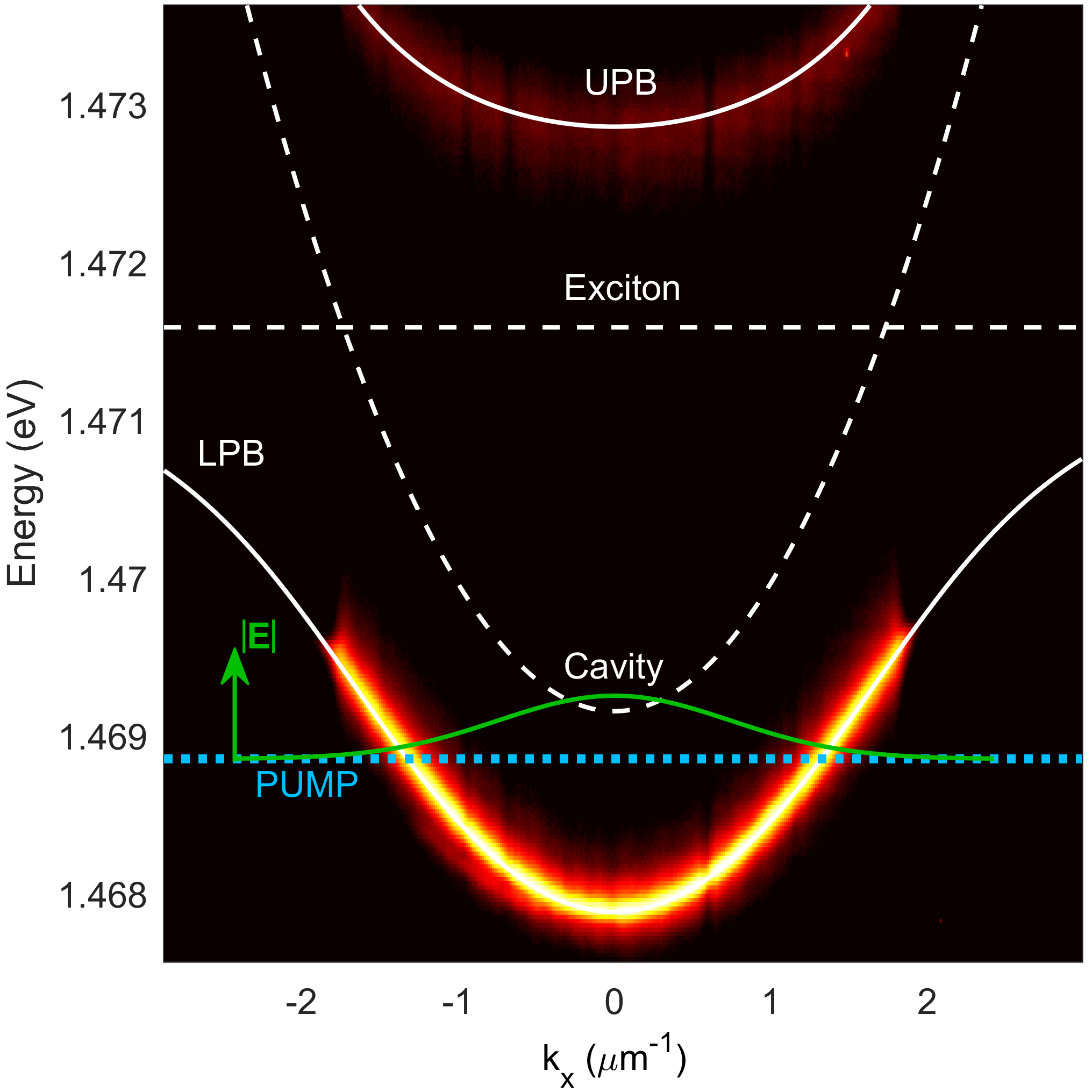}
\center
\caption{\label{fig1} \textbf{Pumping scheme}. Experimentally measured angle-resolved photoluminescence spectrum with fitted curves (white lines) showing both the uncoupled (dashed) and coupled (solid) modes of the system. The blue dotted line shows the energy of the pump laser and the green solid line shows its Gaussian intensity profile and accompanying axis. $|E|$ is the electric field amplitude.}
\end{figure}

Our study utilizes a $\lambda$/2 GaAs cavity, containing three 10 nm In$_{0.04}$Ga$_{0.96}$As quantum wells (QWs), layered between two Al$_{0.85}$Ga$_{0.15}$As/GaAs distributed Bragg reflectors with 23 (27) top (bottom) pairs. The cavity-exciton detuning is approximately -2.5 meV and the strong coupling of the cavity mode and QW excitons results in a Rabi splitting of about 4.5 meV. Our sample is mounted in a cold finger cryostat kept at 5 K, with large angular optical access in both excitation and detection paths. We work in transmission geometry, exciting the microcavity at normal incidence with 100 ps pulses from a tunable mode-locked Ti:Sapphire laser with an 80 MHz repetition rate. The beam is focused to a 2.5 $\mu$m spot (we varied the spot size to produce different patterns) using a microscope objective with a numerical aperture of 0.42; a second microscope objective collects the emitted light, which is then focused onto the entrance slit of an imaging spectrometer connected to a CCD camera to record images.

Our system is pumped with a right-circularly polarized beam that is blue-detuned by between 0.5 and 1 meV from the bottom of the lower polariton branch. A schematic of the excitation scheme can be seen in Fig. 1. In this configuration the pump excites polariton plane waves on a ring in momentum space with fixed magnitude of the in-plane wavevector, $|\textbf{k}|$, since the tight focusing of the beam in real space leads to a broad profile in Fourier space where the Gaussian wings directly excite the polariton dispersion (see Fig. 1). As a result, at low pumping powers the real space polariton density distribution is well approximated by a zeroth order Bessel function of the first kind (Fig. 2b) \cite{Dominici2015}, which is the Fourier transform of a ring. Thus the separation of the Bessel rings is determined by the radius of the ring in momentum space. Note that some weak modulation of the low power polariton density distribution is observed in Fig. 2b.  We believe that the origin of this structure is anisotropy along the crystallographic axes which weakly modulates the cavity transmission, leading to a directionally dependent transmission intensity.

In Fig. 2a we probe the response of the intracavity population against pumping power by plotting the total intensity of the light emitted by the microcavity. In the low power regime, the system behaves linearly with pumping power. At around 3.5 mW (quoted powers are time-averaged and measured immediately before the excitation objective) we observe a threshold which we denote as $P_{1}$. At this point there is a change in the slope of the curve, which corresponds physically to the dynamical instability point where the azimuthal symmetry of the intracavity field is spontaneously broken and nontrivial steady-state patterns become possible. A selection of the clearest patterns is shown in Figure 2c--g, the first of which is a dipole state with two bright lobes on the innermost ring surrounding the central spot. As the pumping power is increased, the steady state evolves to accommodate four lobes on the innermost ring, replacing the dipole state (Fig. 2d). Beyond this pumping power there is a second threshold denoted as $P_{2}$, where there is an abrupt discontinuity in the emission intensity as the intracavity field jumps to a new stable branch (Fig. 2a). In this regime we see an enlarged central spot with more azimuthal polygon patterns observed at longer radii. Here we show hexagon, heptagon and octagon pattern states (Fig. 2e, f and g respectively). At even higher pumping powers the system enters a turbulent regime where the coexistence of a large number of modes signals the onset of spatial turbulence \footnote{See Supplemental Material}. 

\begin{figure}
\center
\includegraphics[scale=1]{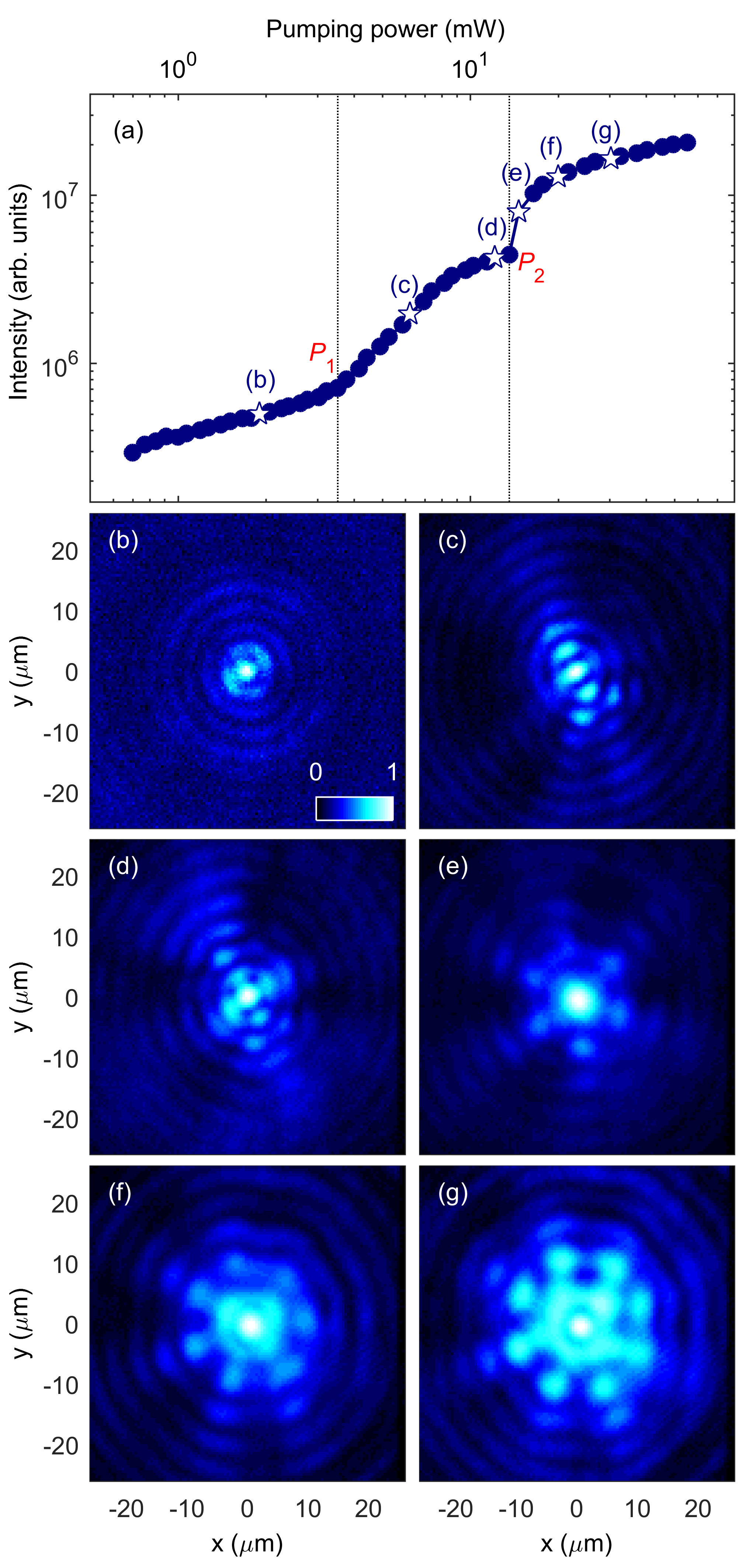}
\center
\caption{\label{fig2} \textbf{Pump power dependence}. (a) Integrated  emission intensity as a function of pump power. The unfilled stars are pumping powers corresponding to the real-space density distributions shown in (b)-(g). $P_{1}$ and $P_{2}$ mark the onset of parametric scattering and bistability respectively. (b) At low pumping powers we observe a Bessel distribution of polaritons. Above $P_{1}$ we form patterns with two (c) and four (d) bright lobes. Above $P_{2}$ patterns with six (e), seven (f) and eight (g) bright lobes emerge. The data are plotted on a log scale and normalized.}
\end{figure}

We now explain the mechanisms responsible for pattern formation. With increasing pump power polariton-polariton scattering starts playing an important role and above the $P_1$ threshold the system becomes unstable against the growth of small perturbations in the form of noise fluctuations, as certain transverse modes begin to experience a gain  that overcomes losses. The polariton repulsive nonlinearity also blueshifts the LPB upwards in energy (an effect occurring most strongly at the centre of the pump spot) and scattering channels to new transverse modes open up. This allows macroscopic populations of particles to accumulate in nontrivial pattern states, spontaneously breaking azimuthal symmetry, as pumped polaritons elastically scatter to signal and idler modes at the same frequency (energy-degenerate parametric scattering) and the system enters what can be described as an OPO regime. Both the blueshift of LPB upwards and parametric scattering  results in the change of slope of the power dependence curve above threshold $P_1$ seen in Fig. 2a because i) the excitation of polaritons on a ring in momentum space, which decreases with pump power, becomes more efficiently pumped by the Gaussian pump and ii) the transfer of polaritons into signal and idler modes increases the energy of the total polariton field inside the cavity.

In contrast to conventional polariton OPOs \cite{PhysRevB.62.R16247,PhysRevLett.85.3680,PhysRevB.62.R13298,PhysRevB.77.115336}, where pump, signal and idler modes are plane waves with well defined $\textbf{k}$  and different frequencies, the pump state in our case is a coherent superposition of polariton plane waves on a ring, and parametrically generated signal and idler states consist of a coherent superposition of harmonics carrying OAM of opposite sign and residing on the same ring in Fourier space (see Fig. 3 and Sec IV for experimental and theoretical Fourier space images of patterns). Signal and idler states with non-zero OAM have been observed in both theory \cite{Whittaker2007,Hamp2015} and experiment in conventional OPOs using an external seed carrying OAM \cite{PhysRevLett.104.126402}, and quantized OAM has been shown to be conserved during four wave mixing processes \cite{PhysRevLett.90.133001,Pyragaite2001459}. The dynamics of vortex-antivortex pairs which form in OPO condensates perturbed resonantly with a Gaussian probe beam have been also investigated \cite{Anton:12}. In our system, these vortex states arise spontaneously without external seeding, and the interference between these vortex/antivortex pairs creates a standing wave with $2m$ lobes, where $m$ represents the winding number of signal and idler modes, which must be equal and opposite. It is then the coherent superposition of this standing wave with the pump field with constant phase which gives rise to stable patterns with $m$ lobes, since alternating lobes experience either constructive or destructive interference with the pump. Such an interference effect is pronounced because of the comparable populations of signal and idler states with the pump, due to highly efficient polariton-polariton scattering. This mechanism is validated by the theoretical model used in Sec. IV, and explains how patterns with arbitrary (even or odd) numbers of lobes can arise without violating any conservation laws.

The allowed structures in a pattern forming system are determined by the nonlinear dynamical instabilities present. Generally, instabilities in fluid dynamics, nonlinear optics, chemical reactions, excitable biological media etc. favor patterns such as rolls, stripes, spots and spirals which have been studied extensively \cite{cross,Arecchi1999}. However, the instabilities driving pattern formation are azimuthal in our polariton system which explains pattern geometries (odd polygons) which have not to the best of our knowledge been observed elsewhere.  

In our system, parametric gain is maximized at local maxima of the pump field intensity distribution, i.e. in the center and on the first and second Bessel rings, where the pump intensity is the highest.  In this case the winding number $\pm{m}$ of vortex/antivortex pairs (and hence the number of bright lobes in the final pattern) is strictly defined by the phase-matching condition for quantization of OAM $2m\pi$ = $2\pi{r}k_{\phi}$, where $r$ is the radius of the Bessel rings, which is given by the wave vector $k$ of resonantly injected polaritons (whose wavevector is purely radial), and $k_{\phi}$ corresponds to the azimuthal component of the wavevector of signal and idler polaritons acquired during scattering process. Since the LPB blueshifts as a function of the driving intensity of the pump field, the wavevector $|\textbf{k}|$ of injected polaritons and hence the radii of Bessel rings depends on the pump power. In addition there is a radial dependence of the LPB blueshift created by the inhomogeneous Gaussian profile of the pump, providing the necessary conditions for the formation of various patterns, governed by the pump detuning and polariton dispersion and independent of the system size, characterized by different winding numbers $m$, and with different values of $r$ which correspond to the value of $|\textbf{k}|$ of the patterns. In accordance with this mechanism, we see that $m$ and $r$ increase with pump power due to blueshift of the polariton dispersion and the decreasing $|\textbf{k}|$ of pattern states (see Fig. 2, Fig. 3 and Sec. IV for experimental and theoretical verification). Also qualitatively the increase in the number of lobes with driving intensity can be explained by increased scattering at higher particle densities, such that more of the radial momentum of injected polaritons is converted to OAM.

The second threshold in the system results from bistability of the coherently driven intracavity field, which is characterized by an S-shaped curve connecting a lower and upper stable branch (see Sec. IV). The positive feedback between the field intensity and the blueshift energy means that the pumping efficiency rapidly increases as the LPB ground state approaches the pump energy, leading to a superlinear increase in the emission intensity as the intracavity field jumps to the upper bistable branch at the $P_{2}$ threshold (Fig. 2a). We observe that this is accompanied by the generation of super-Poissonian light and a huge increase in the signal noise in the window of pumping powers where more than one stable solution exists (see Sec. III). 

\begin{figure}
\center
\includegraphics[scale=1]{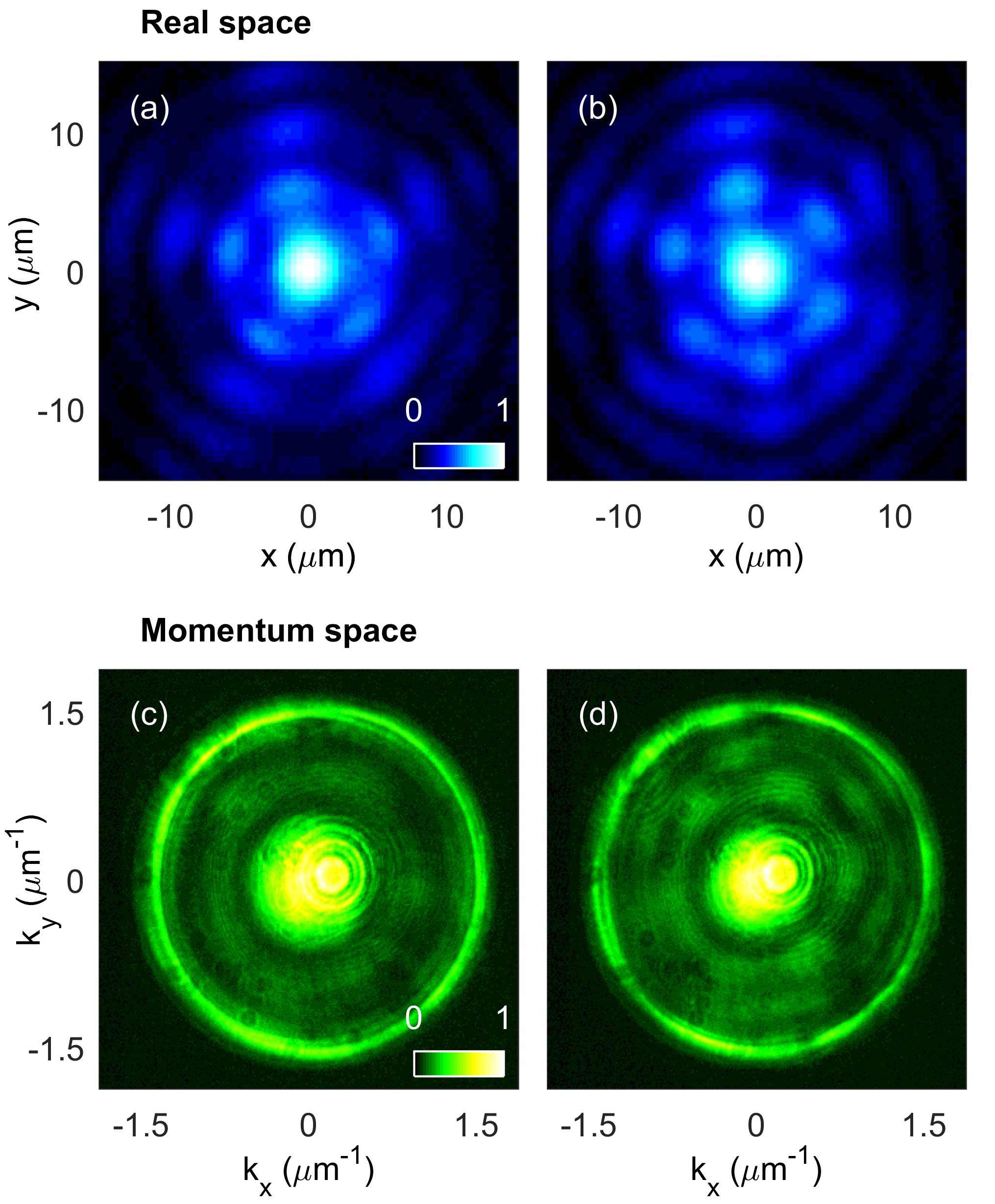}
\center
\caption{\label{fig3} \textbf{Real and momentum space patterns}. Moving the beam waist position relative to the sample surface by adjusting the focus of the excitation objective allows the transformation from a pentagon (a,c) to a hexagon (b,d). Bright lobes can be seen in both real space (upper) and momentum space (lower).}
\end{figure}

By varying the beam profile under fixed driving intensity, it is possible to switch between different stable patterns, achieving a transition between states with different numbers of lobes. An example of this is shown in Fig. 3, where adjusting the size of the pump beam by translating the excitation objective changes the incident power density. This enables the transition from a pentagon to a hexagon, indicating how the most favorable pattern state is highly sensitive to pumping conditions. In addition to the stationary patterns observed in the real space density distribution (Fig. 3a and b), accompanying Fourier space images also reveal polygon-type multi-lobed patterns (Fig. 3c and d). Here we see a large emission from $k=0$, reflecting the fact that the pumping power is above the $P_{2}$ threshold and hence the ground state has blueshifted into resonance with the pump. The larger ring simply corresponds to lower density regions of space, i.e. away from the pump spot, where the LPB is unperturbed. It is in between the outer ring and central spot that we see polygon patterns, arising from the directional instabilities (polaritons gain an azimuthal component of polariton wavevector through parametric scattering) experienced during radial propagation from high to low density regions of space. 

Now we address the question of what determines the orientation of the patterns. In the absence of spatial inhomogeneity, noise fluctuations induce some asymmetry into the initial field distribution created under pulsed excitation. The unstable mode then grows exponentially from this noise, pinning the locations of the polygon maxima. The resulting patterns would be expected to form with random spatial orientation from shot to shot, as they only depend on the relative phase of signal and idler vortex/antivortex states, which is free to evolve from the phase of the pump. Since our measurements are made on macroscopic time scales, averaged over millions of pulses, there must be an additional mechanism stabilizing the patterns and pinning the signal/idler phases. In reality sample disorder \cite{Sanvitto2006} provides spatial inhomogeneity which induces weak anisotropy, which can stabilize the spatial orientation of the patterns from shot to shot in experiment. In addition, in experiment it was also found that for some positions on the sample a rotation of the pattern orientation is observed when changing from left to right circularly polarized pumping \cite{Note1}. In other sample positions the rotation was not observed. The weak localized birefringence naturally present in the system will convert circularly polarized light into elliptically polarized light inside the sample, with the direction of ellipticity depending on the circular polarization. We suggest that this birefringence effect along with TE-TM splitting and spin-dependent polariton-polariton scattering \cite{PSSC:PSSC200562002,KLOPOTOWSKI2006511,Renucci2005,Krizhanovskii2006} may also break azimuthal symmetry and hence define the position of polygon maxima in real space for some positions across the sample where the photonic disorder is weak. A detailed investigation of this effect is outside the scope of this paper.

\section{Photon statistics measurements}

Here we investigate the effect of the parametric instability responsible for pattern formation on the statistical properties of light emitted by the microcavity under coherent driving, by measuring the second order correlation function $g^{(2)}$. For this we used a Hanbury-Brown Twiss interferometer setup with two avalanche photodiodes with a time resolution of 100 ps. Signals from the detectors were recorded as a histogram of coincidences at different time delays, using a timing card with a 24 ps time-bin width. The expression for number of coincidences $G^{(2)}$ is given by
\begin{equation}
G^{(2)}(\tau)=\left \langle {n_{1}(t)n_{2}(t+\tau)} \right \rangle
\end{equation}
where \({n_{1}}\) and \({n_{2}}\) are the numbers of photons registered on the first and second detectors respectively at time $t$ and $t+\tau$ and the angle brackets represent averaging over time $t$. 

In our pulsed excitation scheme, the raw data corresponding to photon coincidences consists of multiple peaks separated by a delay $T$=12.5 ns, which is the repetition rate of the driving laser (see inset of Fig. 4). The number of coincidences $G^{(2)}(j)$ between photons in pulses separated by a time $jT$ (where $j$ is an integer) can be obtained by integrating $G^{(2)}(\tau + jT)$ over a range of $\tau$ to obtain the pulse areas \cite{Deng199}. Assuming that there is no correlation between intensities of consecutive pulses, $G^{(2)}(j)$ may be normalized to the average area of the uncorrelated peaks at $j\neq 0$. The normalized $g^{(2)}(j)$ is then given explicitly by the formula: 
\begin{equation}
g^{(2)}(j)= \frac{\int_{-\Delta t/2}^{+\Delta t/2} \left \langle {n_{1}(t)n_{2}(t+\tau+jT)} \right \rangle d\tau} {\frac{1}{N-1} \sum\limits_{i\neq 0} \int_{-\Delta t/2}^{+\Delta t/2} \left \langle {n_{1}(t)} {n_{2}(t+\tau+iT)} \right \rangle d\tau} 
\end{equation}
\noindent
where $N$=5 is the number of pulses recorded. The integration range was chosen $\Delta t$=6.5 ns, which is roughly half the value of $T$.

The power dependence of $g^{(2)}(0)$ is plotted in Fig. 5a. At pumping powers below the $P_{2}$ threshold \footnote{Data in Fig. 5 were taken for a linearly polarized pump, for which the strength of polariton-polariton interactions is about twice smaller than in the case of circularly polarized pump. This leads to thresholds $P_{1}$ and $P_{2}$ higher than in Fig. 2.}, $g^{(2)}(0)$ values remain at a value of 1 representing a coherent state. At the onset of the $P_{2}$ threshold,  the $g^{(2)}(0)$ value suddenly increases to 1.75 (see Fig. 4 for the $g^{(2)}$ profile of individual signal pulses) before sharply returning to 1 with further increase of pumping power above threshold. We note that despite the narrow resonance of $g^{(2)}(0)$, these results are very reproducible from day to day measurements.

\begin{figure}
\center
\includegraphics[scale=1]{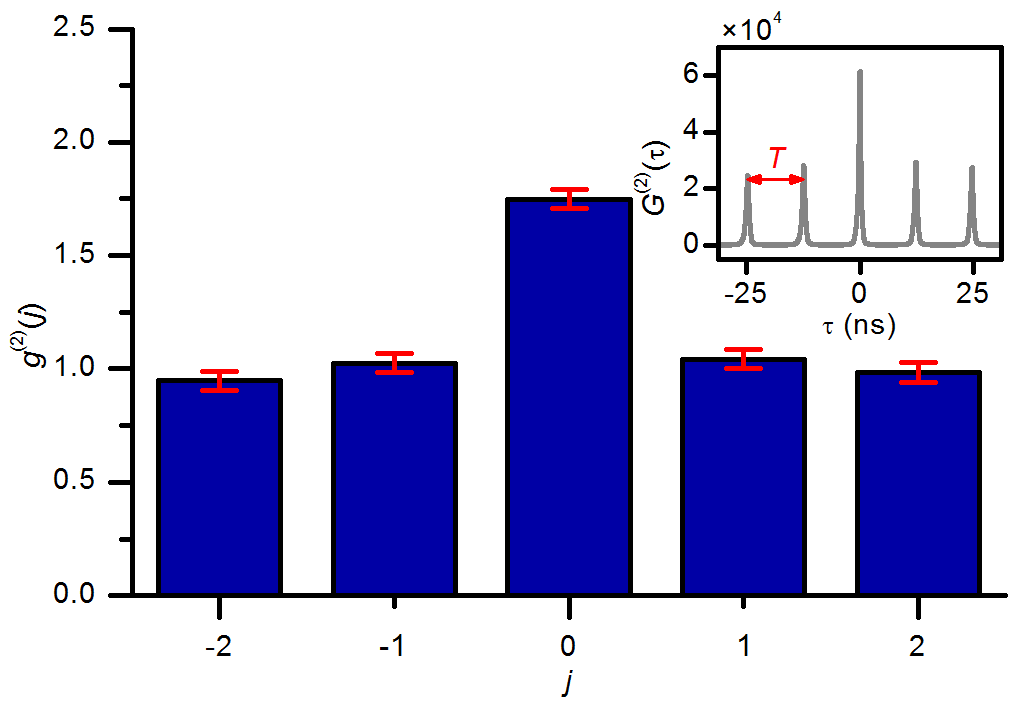}
\center
\caption{\label{fig4} \textbf{Value of $g^{(2)}$ across sequence of five pulses}. At the bistable threshold the value of $g^{(2)}(0)$ shows a bunched value of 1.75  (see Fig 5a). The corresponding raw data consists of photon coincidences at the HBT photodetectors and is shown in the inset.}
\end{figure}

\begin{figure}
\center
\includegraphics[scale=1]{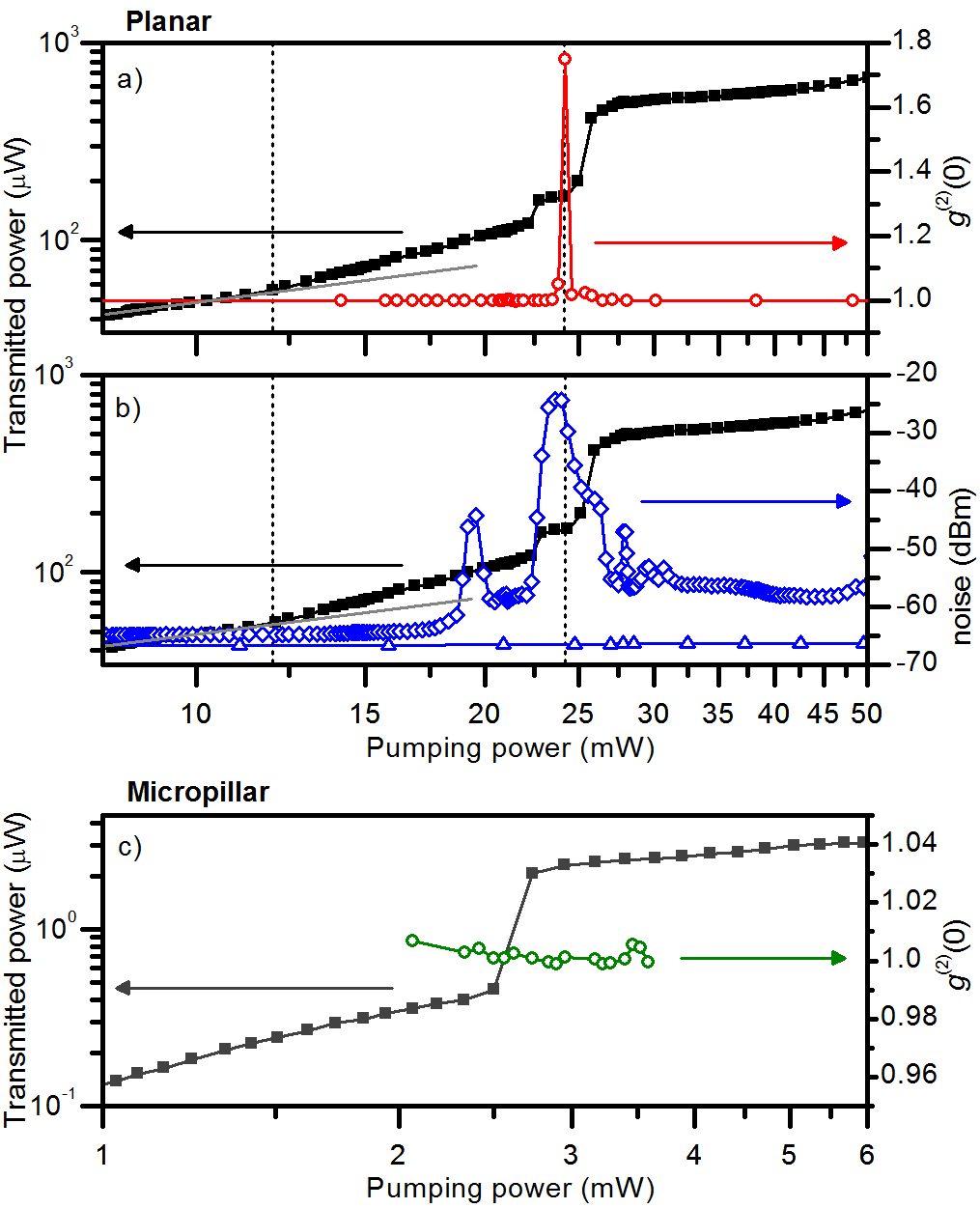}
\center
\caption{\label{fig7} \textbf{Power dependence of $g^{(2)}(0)$ and signal noise}. (a) The transmitted power through our planar cavity sample (black squares) and the corresponding value of $g^{(2)}(0)$ (red circles) against pumping power. (b) The measured photocurrent noise of the emitted light (blue diamonds) shown with the shot noise level (blue triangles) and the transmitted power (black squares) against pumping power. The vertical dotted lines show the two $P_{1}$ and $P_{2}$ thresholds. (c) The value of $g^{(2)}(0)$ (green circles) along with transmitted power (gray squares) against pumping power in the micropillar sample.}
\end{figure}

In addition to the results obtained in our planar cavity sample, we also measured photon correlations in a micropillar. This represents a similar system to our planar cavity but differing in the key aspect of lateral confinement. Parametric scattering is suppressed by the absence of a continuum of transverse modes to which polaritons can scatter, which also precludes pattern formation. A micropillar then represents a system with pure bistability. By selecting the same pumping conditions as for the planar cavity case, we are able to investigate the dynamics of a bistable polariton system without inter-mode scattering. We find that in our micropillar the value of $g^{(2)}(0)$ remains at that of a coherent state value across the range of pumping powers (See Fig. 5c). Such observation indicates that the presence of parametric scattering clearly plays a significant role in the bunching effect in planar structures. 

Let us now discuss the results of our $g^{(2)}(0)$ correlation measurements. In the planar cavity case, a value close to unity measured below the $P_{1}$ threshold is expected since polaritons are quasiresonantly created by laser pulses, inheriting their coherence properties. In this regime the pump field passes through the cavity with minimal nonlinear interactions. Even above the threshold for parametric scattering, the light generated by four wave mixing processes is in a coherent state, and the statistics of the microcavity emission remain firmly Poissonian. It is only when the bistable threshold is reached that intensity fluctuations are strongly altered with the microcavity light emission exhibiting super-Poissonian statistics. 

It has been shown in a multi-mode bistable polariton system that energy can gradually accumulate with time (within tens of picoseconds) in modes populated through parametric scattering, followed by a sharp transition from the lower to the upper steady state branch when the internal energy of the system is sufficient \cite{PhysRevB.90.205303}. We suggest that a similar process is responsible for the bunching effect observed in our system, such that signal and idler modes accumulate particles through the parametric instability responsible for pattern formation within the duration of the pump pulse, before a strongly superlinear increase (in time) of pump, signal and idler fields launches the system onto the upper bistable branch (so-called "blowup" temporal dynamics). In the vicinity of this threshold the populations of pump, signal and idler modes integrated over the pump pulse duration are highly sensitive to initial conditions, and thus to quantum fluctuations of the pump and photonic field inside the cavity. This leads to strong noise in the cavity emission from pulse to pulse and thus to the super-Poissonian photon bunching observed experimentally. By contrast, in a single micropillar cavity there is no superlinear temporal dynamics of the intracavity field due to the absence of parametric scattering channels to transverse modes, and before the bistability threshold is reached (the right turning point of the S-shaped curve) the value of quantum fluctuations is insufficient to drive the system into stochastic resonance, precluding the observation of the strong bunching effect. 

Above the $P_{2}$ threshold, the system resides on the upper bistable branch with a well defined polariton population integrated over each pump pulse, so the value of $g^{(2)}(0)$ is restored to unity, and the statistics revert back to Poissonian. The behavior in our system differs from that of non-resonantly pumped polariton lasers and condensates \cite{PhysRevLett.100.067402,:/content/aip/journal/apl/107/22/10.1063/1.4936889,PhysRevB.85.075318} where photon bunching is attributed to thermal populations of polaritons, which subsequently scatter into the coherent ground state at higher intensities or longer times in the case of above-threshold pumping. 

Simultaneously, we measured the photocurrent power noise arising from the microcavity emission. This allows us to directly probe the statistics of the emission, so it is complementary to the measurements of $g^{(2)}$ \cite{Bachor2004}. The polariton signal was sent to a Si photodiode detector, generating a photocurrent which is amplified and its power noise is analyzed in  the frequency domain by a spectrum analyzer. The photocurrent is acquired in a 1 MHz bandwidth around a central frequency of 5 MHz, well separated from the 80 MHz repetition rate of the laser.

The power dependence of the noise signal is shown in Fig. 5b. At low pumping powers $<10$ mW, noise levels remain relatively close to shot noise levels, i.e. noise expected from purely Poissonian fluctuations, depicted by unfilled triangles. An increase of noise by a few dB above shot noise level appears at pumping power $\sim 15$ mW before the bistability threshold $P_{2}\sim 25$ mW. This is followed by a further sharp increase (20 dB) in noise at $\sim$18 mW and then at $\sim$22 mW pumping powers, producing a wide peak with a shoulder at a lower, but still elevated noise level, in the vicinity of the $P_{2}\sim 25$ mW threshold.  The noise reduces at higher powers, but oscillations in the noise signal are still seen. The underlying physics differs from the power dependence of $g^{(2)}(0)$, which reflects amplification of high frequency (80 MHz) quantum fluctuations, i.e. pulse-to-pulse photon number variations. By contrast, low frequency (5 MHz) noise is likely to originate from amplified classical noise of the pump field, which grows quadratically with excitation power and is probably much stronger than the quantum noise of the pump which depends linearly on excitation power \cite{Bachor2004}. We believe that the noise amplification occurs through generation of fluctuating signal and idler modes and possibly stochastic jumps between lower and upper bistable polariton states similarly to the mechanisms responsible for the enhanced  $g^{(2)}(0)$, but this may happen over a broader interval of pump energies since classical fluctuations are stronger than quantum. Different maxima in the power dependence of noise are probably governed by the parametric instabilities responsible for pattern formation below and above the bistability threshold $P_{2}$ as observed in Fig. 2. It is known in other nonlinear pattern forming systems that noise may be amplified at pump parameters where there is crossover between different pattern states, referred to as domain coexistence \cite{PhysRevLett.76.1063}. In that case the noise originates from two competing unstable modes both experiencing positive growth rates, very similar to the situation we observe here (see Sec IV).

\section{Theoretical analysis}

We now present a theoretical analysis which provides additional insight into the mechanisms underlying pattern formation, substantiating further the interpretations presented in section II. We perform single-shot numerical simulations of the propagation of a pump beam through our cavity in the presence of symmetry-breaking perturbations. We begin with the widely accepted mean-field model for excitons strongly coupled to circularly polarized cavity photons \cite{sanvitto, kavokin, flayac, manni, baas} whose fields are given as follows:

\begin{equation}\begin{split}
\label{eq:secor}
&\partial _t {E^ \pm } - i\nabla _ \bot ^2{E^ \pm } + \left[ {\gamma  - i\left( {{\omega _p} + \delta } \right)} \right]{E^ \pm } \\& = i{\Omega _R}{\Psi ^ \pm } + E_p^ \pm (x,y),
\end{split}
\end{equation}

\begin{equation}\begin{split}
\label{eq:secor}
\partial _t{\Psi ^ \pm } + \left[ {\gamma  - i{\omega _p}} \right]{\Psi ^ \pm } &+ i\left( {{{\left| {{\Psi ^ \pm }} \right|}^2} + \alpha {{\left| {{\Psi ^ \mp }} \right|}^2}} \right){\Psi ^ \pm } \\& = i{\Omega _R}{E^ \pm }
\end{split}
\end{equation}

Here $E^{\pm}$ and $\Psi^{\pm}$ are the complex amplitudes of the photonic field and coherent excitons obtained through a standard averaging procedure of the related creation or annihilation operators. Normalization is such that $g^{-1}\lvert{E^{\pm}}\rvert^{2}$ and $g^{-1}\lvert{\Psi^{\pm}}\rvert^{2}$ are the photon and exciton numbers per unit area. Here $g$ is the exciton-exciton interaction constant. The symbols + and - represent right and left circular polarization of light and the corresponding exciton spins. $\gamma$ denotes the cavity and exciton damping constants. $\omega_{p} = \omega - \omega_{0}$, $\delta = \omega_{0} - \omega_{c}$ describe the detunings of the pump frequency $\omega$ and the cavity $\omega_{c}$ from the excitonic resonance $\omega_{0}$. $\Omega_{R}$ is the Rabi frequency. The two spin components of the excitons are coupled by the dimensionless parameter $\alpha$ which exhibit values around $\alpha$ = -0.1 \cite{sanvitto, kavokin, flayac}. 

We start our analysis, which begins at the onset of the $P_{1}$ threshold, with a circularly polarized coherent pump which has an azimuthally-symmetric Gaussian profile:

\begin{equation}
\label{eq:secor}
E_p^ + (x,y) = A{}_0{e^{ - {r^2}({R^{ - 2}} + i\eta )}},
\end{equation}

\begin{equation}
\label{eq:secor}
E_p^ - (x,y) = 0
\end{equation}

\noindent
where $r=\sqrt{x^{2} + y^{2}}$. A finite phase curvature of the pump beam is characterized by the constant $\eta$ and reflects the fact that in experiment the beam waist was displaced with respect to the position of the quantum wells, resulting from the small adjustments of the excitation objective from its focal distance used to produce the most pronounced patterns. A non-zero and negative $\eta$ is used in our theoretical analysis to provide a good qualitative agreement with experiment, although the phase curvature is not fundamental to the underlying nonlinear processes themselves so patterns are still observable in both experiment and theory over a range of values of $\eta$. 
 
\begin{figure}
\center
\includegraphics[scale=0.3]{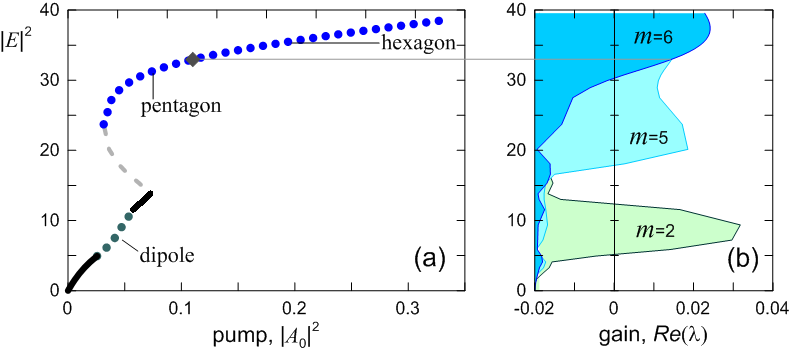}
\center
\caption{\label{fig8} \textbf{Pump power dependence and linear stability analysis}. (a) Maximum intensity of radially-symmetric steady-state against pump power. The pump beam has a Gaussian profile with radius R=18 $\mu$m and phase curvature $\eta=-0.02$ $\mu$m$^{-2}$. Stable states are depicted by solid lines whereas dotted lines show unstable solutions breaking spontaneously their radial symmetry. The dashed line represents solutions which are unstable against $m=0$ perturbations. (b) Growth rates of leading unstable modes (with angular numbers $m$=2, 5, 6) against intensity of the radially-symmetric solution. Other parameters: $\delta$=-2.561 meV, $\omega_{p}$=-3.3 meV, $\gamma$=0.02 meV, $\Omega_{R}$=2.142 meV. These parameters are very close to the experimental conditions of Fig. 2.}
\end{figure}

First, we consider the radially-symmetric steady-state response which is independent of the polar angle $\theta$=arg($x + iy$). Applying the Newton iterative method, we have calculated radially-symmetric steady-states for different driving intensities (see Fig. 6a). In agreement with pioneering works on exciton-polariton dynamics \cite{baas}, the obtained solutions are bistable provided the pump frequency is blue-detuned by more than a certain amount with respect to the LPB. The threshold $P_{2}$ corresponding to the abrupt intensity transition in the experiment (Fig. 2a) is expected to occur at the turning point of the lower bistability branch, approximately where the pump equals 0.075 in Fig. 6a. In the limit of weak pumping, the spatial intensity profiles of the intracavity field possess the previously described ring structure (see Fig. 7a). This confirms that the pump beam excites polaritons on the LPB which are in resonance with the pump frequency most strongly, compared to the $k=0$ mode, which is comparatively very weakly driven (see Fig. 7b).

\begin{figure}
\center
\includegraphics[scale=1]{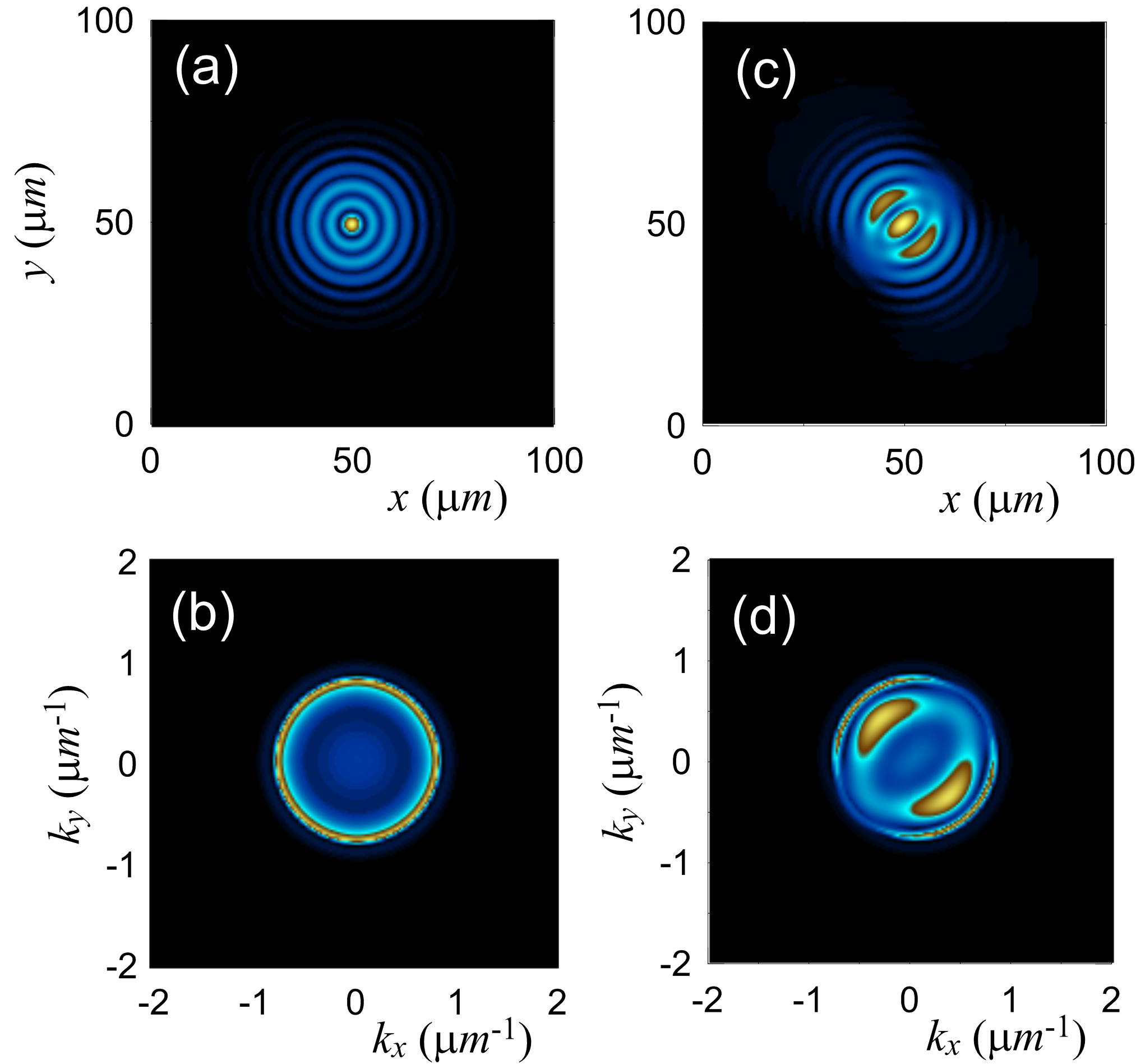}
\center
\caption{\label{fig9} \textbf{Calculated polariton density distributions below bistability threshold}. Intensity profiles of steady-state solutions are shown in real (a, c) and momentum (b, d) space. (a, b) Radially-symmetric solution in a low-pump limit for $A_0^{2}$=0.01. (c, d) Stable dipole state for $A_0^{2}$=0.0529. Other parameters are similar to those of Fig. 6.}
\end{figure}

We also perform a full two-dimensional linear stability analysis of the localized structures, which provides insight into the pattern selection process by showing which transverse modes experience parametric gain. The linear perturbations are assumed to be vortices carrying OAM with the general form

\begin{equation}
\label{eq:secor}
{a_ + }(r){e^{im{\kern 1pt} {\kern 1pt} \theta  + {\lambda ^{}}(m)t}} + a_ - ^*(r){e^{ - im{\kern 1pt} {\kern 1pt} \theta  + {\lambda ^*}t}}
\end{equation}

\noindent
where the azimuthal winding number is integer, $m$=0, 1, 2, 3... . The solution becomes unstable if at least one of the obtained eigenmodes possesses a positive growth rate, i.e. Re$[\lambda]>0$. The linear stability analysis shows that the radially-symmetric solutions can become unstable with respect to perturbations breaking the azimuthal symmetry, i.e. with $m \neq 0$ , shown in Fig. 6b.
For instance, the linear eigenmode with $m$=2  can destabilize the steady-state field profile for pump intensities below the bistability threshold (see Figs. 6a and b). Direct numerical simulations within the original equations show that this instability develops with time and results in the formation of a dipole-like steady-state with three intensity peaks (see Figs. 7c and 7d) in strong qualitative agreement with the experimental observation of Fig. 2c.

\begin{figure}
\center
\includegraphics[scale=1]{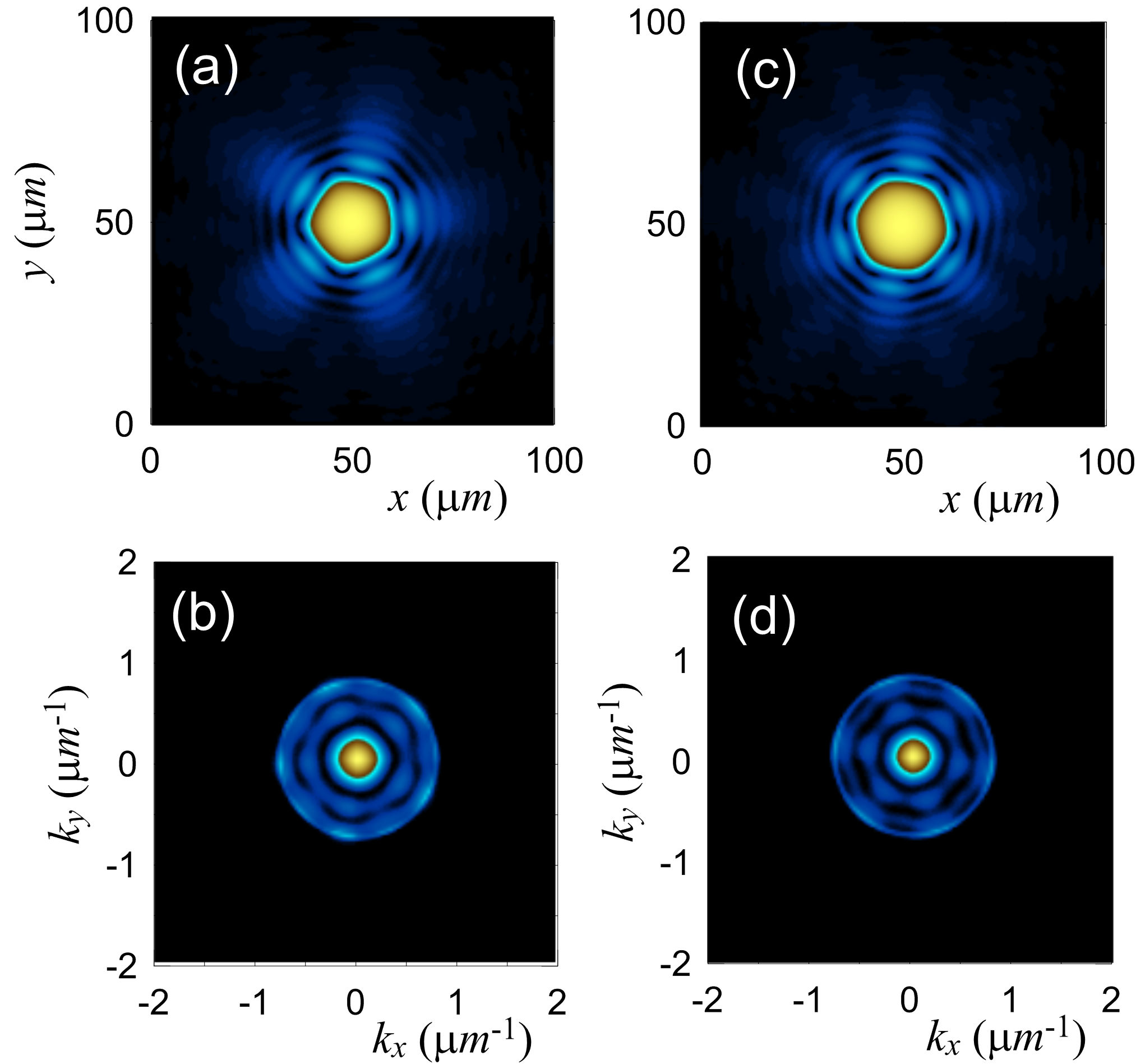}
\center
\caption{\label{fig10} \textbf{Calculated polariton density distributions above bistability threshold}. Intensity profiles of steady-state solutions are shown in real (a,c) and momentum (b,d) space. (a,b) Pentagon state for $A_0^{2}$=0.09. (c,d) Hexagon state for $A_0^{2}$=0.16. Other parameters are similar to those of Fig. 6.}
\end{figure} 

For stronger driving intensities above the bistability threshold, the radially symmetric solutions undergo azimuthal instabilities with larger winding numbers $m$. For instance the growth of the unstable mode with winding number $m$=5 results in the formation of the pentagonal structures shown in Figs. 8a and b. At even higher pumping intensities, the growth rate of the unstable mode with $m$=6 overtakes that of the pentagon and thus a hexagon becomes the most favorable solution, winning the mode competition (see Figs. 8c and d). Such a behaviour is in agreement with pentagon and hexagon experimental observation above bistability threshold in Figs. 2 f and 3.  A spatial Fourier analysis (see Figs. 8b and d) shows strong agreement with experimentally measured far-field intensity profiles above the bistability threshold (Figs. 3c and d), showing the strongly populated $k=0$ mode surrounded by a polygon pattern and the sustained ring structure at larger wave number, which itself shows some azimuthal dependence on the intensity. In numerical simulations we find that switching between the pentagon and hexagon states is possible by changing either the radius or the phase curvature of the pump beam. In experiment both of these scenarios correspond to shifting the focus position of the excitation objective, which irradiates the sample with a different cross-section of the input beam. We showed in Fig. 3 that this switching behavior is simple to achieve, as expected by the theoretical model used here. Despite the simplicity of our model, the above stability analysis of the radially symmetric polariton patterns offers very good qualitative agreement with the near- and far-field pattern profiles observed in Figs. 2 and 3, both before and after the $P_{2}$ threshold.  

\section{Conclusions}

In this work we have reported on the spontaneous formation of polygon patterns in a coherently driven dissipative system of polaritons, investigating how the interplay between parametric instability and bistability affects the system dynamics. Our results present a new and simple paradigm for the creation of patterns with novel geometries in unconfined polariton fluids, which does not rely on the use of pumps with nontrivial geometries or phase, or confinement within heterostructures with multiple dispersions. We have shown how the onset of parametric instability causes coherent scattering to states carrying OAM, leading to spontaneous symmetry-breaking and the emergence of stationary structures. The coherent superposition of a driven pump mode with parametrically unstable signal and idler modes has proven in numerical simulations to be able to reproduce the main behaviours of our system, namely the generation, competition and selection of different patterns similar to those observed in experiment. 

We have also demonstrated a bunching effect at the bistable threshold, where the light emitted by the microcavity exhibits super-Poissonian statistics, which we attribute to the underlying strongly superlinear temporal dynamics of the system. Specifically, it is a nonlinear effect in which the intracavity intensity fluctuations are strongly enhanced by the interplay between the pump-induced bistability and the dynamic ``pattern-forming'' parametric instability, which provides scattering channels into which particles can scatter and accumulate. 

In terms of future directions, there is a vast wealth of potential physics to be explored. Microcavities engineered to feature a large TE-TM splitting are expected to generate patterns with rich polarization properties resulting from an in-plane effective magnetic field and spin-anisotropic interactions \cite{PhysRevB.89.235302}, whilst also offering the observation of new topological excitations, such as Skyrmions and half-solitons \cite{PhysRevLett.110.016404}. The presence of polarization multistability \cite{PhysRevLett.98.236401} remains unexplored in the context of pattern formation. Single shot time-resolved measurements will certainly provide invaluable insight into the pattern formation process. Extreme spatio-temporal statistical fluctuations in the value of the intracavity field known as rogue waves are expected, which may play a key role in the transverse mode dynamics \cite{Dudley2014} and so far have yet to be explored in polariton systems. Furthermore, such measurements will shed light on the bunching effect measured at the threshold. Squeezing has already been observed at the bistable threshold in a single-mode micropillar \cite{Boulier2014}, but theoretically predicted antibunching amplified by parametric scattering \cite{PhysRevA.90.063805} is yet to be observed.  

Beyond exciton-polariton systems, we expect our study to open up a new line of research into the formation of patterns driven by high order vortices. Ideal candidates for further exploration of spontaneous vortex patterns are nonlinear optical systems such as cavities with photorefractive crystals and atomic vapours \cite{Arecchi1999}, where pattern formation relies on a dynamical combination of propagation, nonlinearity, gain and loss as in our case. Further afield, in condensed matter systems such as superfluids and atomic Bose-Einstein condensates described by a conservative Gross-Pitaevskii equation, the opportunity arises in light of our work to investigate what exotic patterns may be formed by matter-wave dynamical instabilities in different dimensionalities and trapping geometries \cite{doi:10.1142/S0217984904006809}. For example, modulational instability leading to the generation of new harmonics was recently reported in an atomic condensate, \cite{1703.07502} but it remains to be seen whether components carrying high order vorticity can be created during pattern formation \cite{1703.01411}. Furthermore, studies of quantum effects in the context of pattern formation such as quantum entanglement, squeezing and spatial correlations are still to be explored, and have potential to be stimulated by our work. 

Experimental data supporting this study are openly available from the University of Sheffield repository \footnote{\url{https://doi.org/10.15131/shef.data.5221408}}.

\section*{Acknowledgements}

We acknowledge support by EPSRC Grant EP/J007544 and EP/N031776/1, ERC Advanced Grant EXCIPOL No. 320570 and the Leverhulme Trust Grant No. PRG-2013- 339. S.S.G. acknowledges financial support from RFBR, Grant No. 16-02-01172 and
Volkswagen Foundation Grant No. 90418. We thank Marzena Szymanska and Dmitry Skryabin for careful reading of the manuscript and useful discussions.

\bibliographystyle{ieeetr}
\bibliography{patterns_paper}

\pagebreak
\widetext
\begin{center}
\textbf{\large Supplementary material}
\end{center}
\setcounter{equation}{0}
\setcounter{figure}{0}
\setcounter{table}{0}
\setcounter{page}{1}
\makeatletter
\renewcommand{\theequation}{S\arabic{equation}}
\renewcommand{\thefigure}{S\arabic{figure}}

\section*{Energy-degenerate parametric scattering}

In Fig. S1 we show angle-resolved photoluminescence spectra of the pattern states, proving that the scattered modes are at the same energy as the pump. At low excitation powers, the polaritons reside on the lower polariton branch (LPB) at the energy of the pump [Fig. S1(a)]. As the pumping intensity increases beyond the $P_{1}$ threshold for parametric scattering, the LPB blueshift, which is stronger for positions closer to the center of the pump spot, gives polaritons a smaller wave vector for the same energy [Fig. S1(b)]. The coherent scattering processes described in the main article occur at these wave vectors, such that macroscopically coherent states with well defined orbital angular momentum are generated on a ring in Fourier space. At high excitation powers, far above the bistable threshold $P_{2}$, the $k=0$ mode is blueshifted into resonance with the pump, dominating the spectrum [Fig. S1(c)]. 

\begin{figure}[!htb]
\center
\includegraphics[scale=0.5]{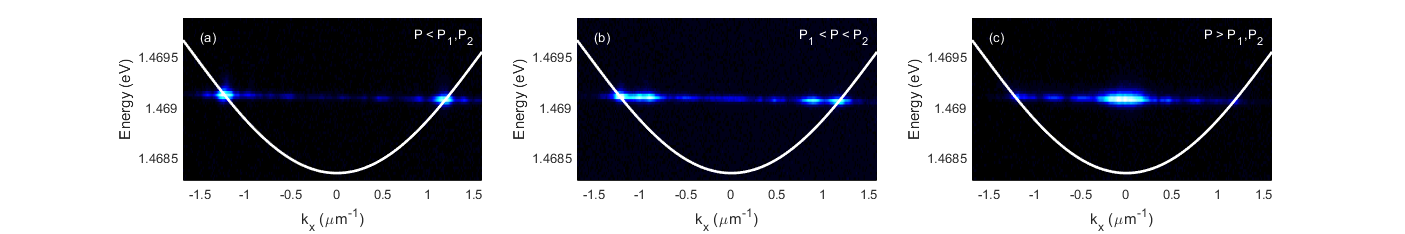}
\center
\caption{\label{fig1} \textbf{Energy-degenerate scattering}. Angle-resolved photoluminescence spectra along $k_{y}=0$ for driving intensities in the three distinct regimes. The white lines show the unperturbed LPB.}
\end{figure}

\section*{Rotation of patterns}

Here we present supplementary measurements on the rotation of patterns, the mechanism of which is discussed in Sec. II of the main text. For certain positions on the sample, changing from left to right circularly polarized pumping allows significant rotation of stable patterns about the central spot. An example is shown in Fig. S2, where rotation of a hexagon state was achieved, such that the polar angle of the polygon maxima changed by a constant offset of about $30\degree$ without an overall change in the form of the pattern. 

\begin{figure}[!htb]
\center
\includegraphics[scale=1]{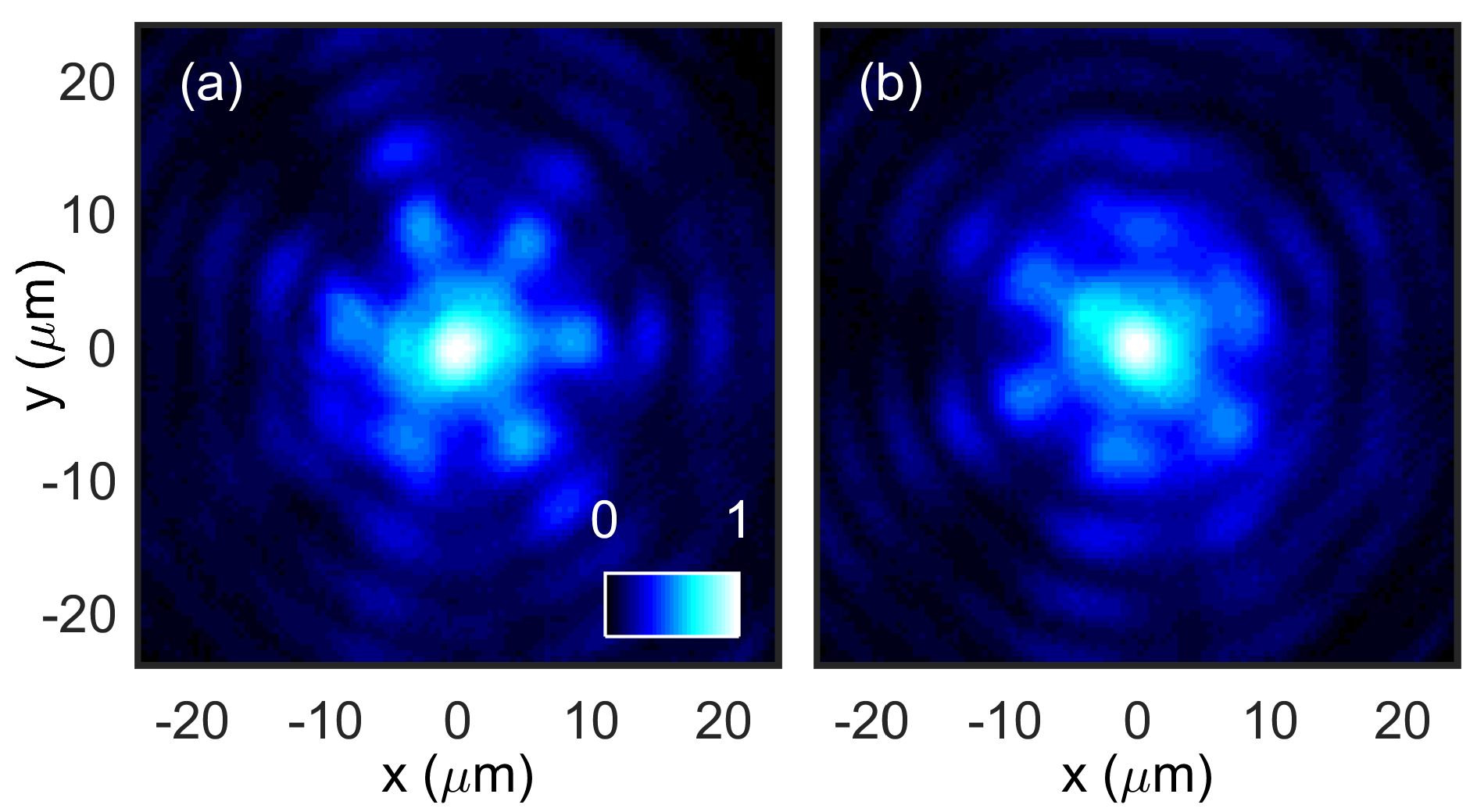}
\center
\caption{\label{fig1} \textbf{Rotation of patterns}. Real space images of stable hexagon patterns. Under a transition from left-(a) to right-(b)circularly polarized pumping the patterns are seen to rotate.}
\end{figure}

In order to clarify the physics behind the polarization-dependent rotation of patterns we performed numerical modeling of the dynamics within the mean-field model discussed in the main text (see Eqs. (3,4)):

\begin{equation}
\label{eq:secor}
\partial _t {E^ \pm } - i\nabla _ \bot ^2{E^ \pm } + \left[ {\gamma  - i\left( {{\omega _p} + \delta } \right)} \right]{E^ \pm } + i\beta\left(\partial _x \mp i\partial_y \right)^{2} {E^ \mp } + i\Delta_{XY} {E^ \mp } = i{\Omega _R}{\Psi ^ \pm } + E_p^ \pm (x,y),
\end{equation}

Additionally to the already discussed parameters (see the main text) we take into account TE-TM splitting of the cavity denoted by $\beta$ and birefringence through the parameter $\Delta_{XY}$. We will demonstrate that the interplay between TE-TM splitting and birefringence can result in the polarization-dependent rotation of pattern states. Figs. S3(a) and S3(b) show a typical example of two real space intensity patterns excited by coherent pump beams with opposite circular polarizations. The orientation of the pattern is defined by a minimum angle between an axis going through pattern maxima and the $x$ axis. Rotation is given by the angle $\alpha^{-}$ between the two axes under excitation with left –circular polarization of the pump beam and angle $\alpha^{+}$ for right –circular polarization (Fig. S3 a, b). In order to elucidate the influence of different physical mechanisms we performed numerical simulation of the total angle of pattern rotation ($\alpha=\alpha^{+}-\alpha^{-}$) depending on the birefringence parameter for two values of TE-TM splitting (Fig. S3c). From fitted curves to the dispersion experimentally measured in TE/TM bases we extract $\beta \sim 0.062$ meV$\mu$m$^{2}$, between the values used in our simulations. It was found that for vanishing birefringence, the patterns are oriented randomly. For small birefringence $<$0.03 meV the patterns are oriented along the $x$ axis and at intermediate values of birefringence (0.03 meV $<\Delta_{XY}<$ 0.06 meV) the pattern axis rotates with respect to the $x$ axis by $\alpha^{+}(\alpha^{-})$ angle for a left(right)-circularly polarized pump. For large birefringence $>$0.07 meV the patterns again tend to orient along the $x$ axis. These numerical simulations show that patterns can be controllably rotated by about 30 degrees by switching between right and left circular polarization of the pump provided the birefringence parameter is strong enough. The theoretical values of $\Delta_{XY}$ at which the rotation of the pattern is observed are consistent with the typical experimental values of birefringence (0.02-0.05 meV) in GaAs microcavity samples. 

\begin{figure}[!htb]
\center
\includegraphics[scale=0.2]{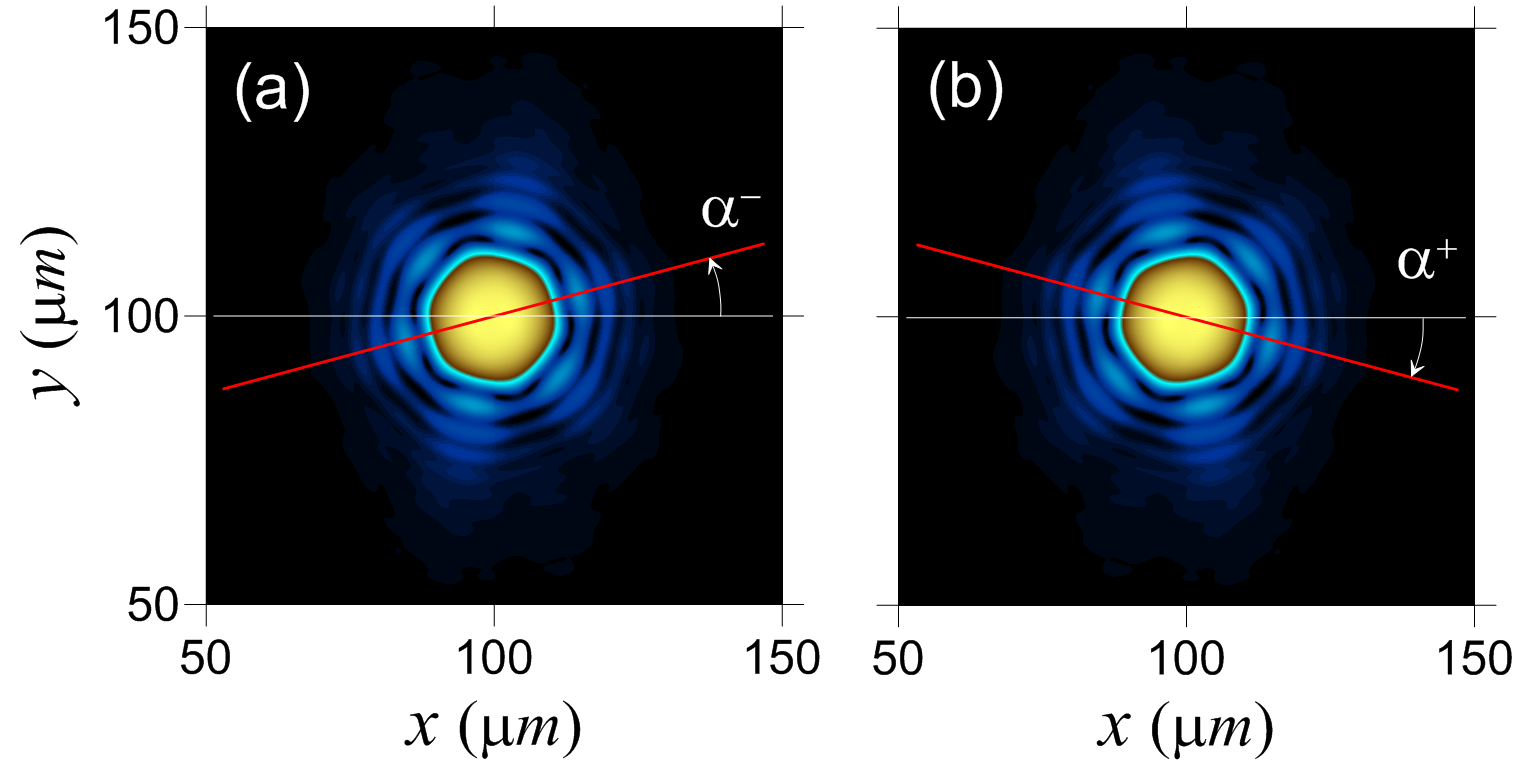}
\includegraphics[scale=0.7]{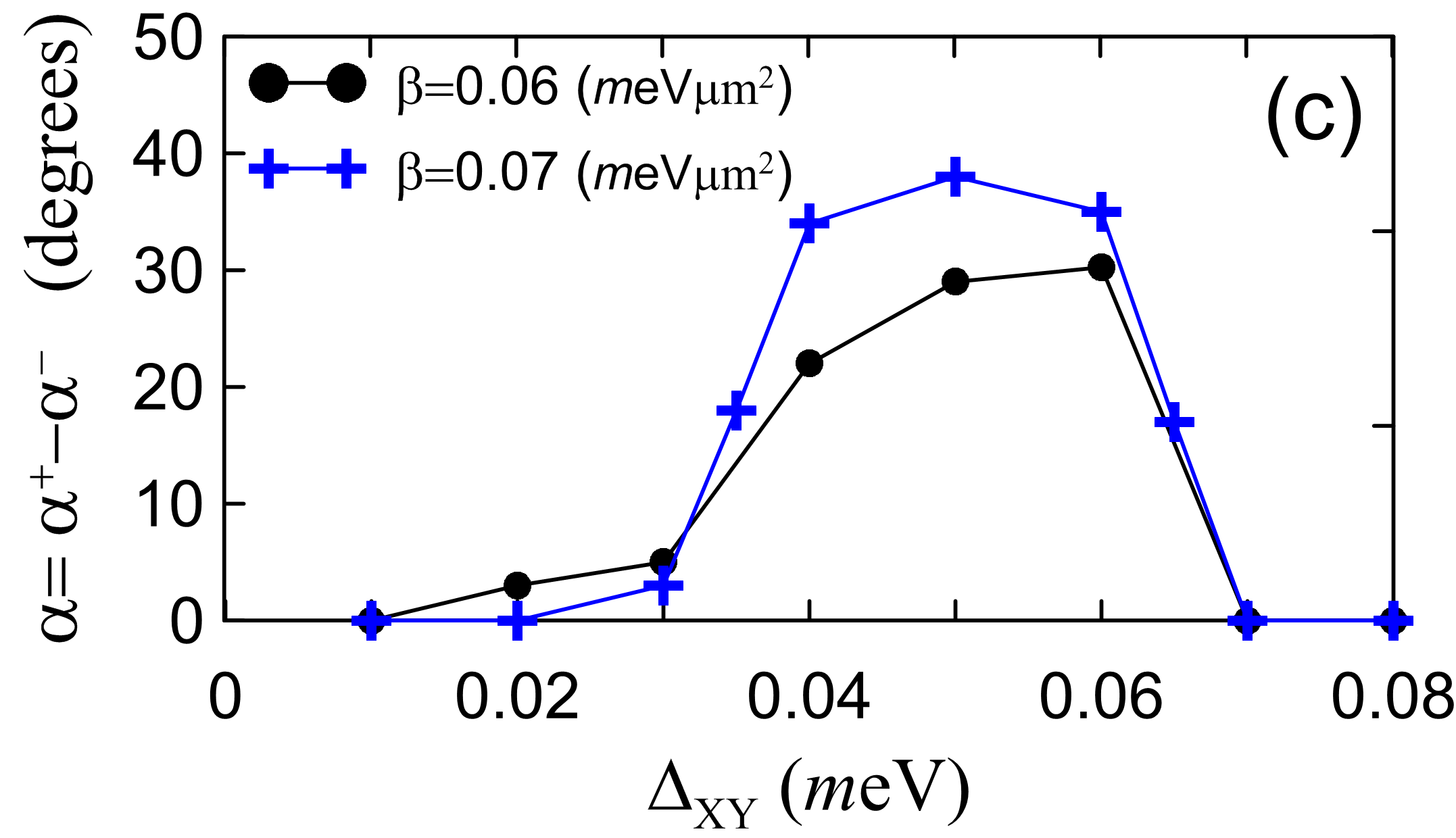}
\center
\caption{\label{fig1} \textbf{Theoretical rotation of patterns}. Calculated intensity profiles of the hexagon patterns under (a) left-circularly-polarized and (b) right-circularly polarized pump beam. Parameters are similar to those of Fig. 8 of the main text. (c) Rotation angle $|\alpha^{+}|+|\alpha^{-}|$  of the pattern vs. the birefringence parameter $\Delta_{XY}$ for two realistic values of the TE-TM splitting.}
\end{figure}

\end{document}